\documentclass[11pt,a4paper,twocolumn]{book}
\usepackage{jnpcs}
\usepackage{amsmath}
\usepackage{graphics}
\usepackage{epsfig}

\newcommand{\be}{\begin{equation}}
\newcommand{\ee}{\end{equation}}
\newcommand{\bn}{\begin{eqnarray}}
\newcommand{\en}{\end{eqnarray}}
\newcommand{\bd}{\begin{displaymath}}
\newcommand{\ed}{\end{displaymath}}
\newcommand{\bnn}{\begin{eqnarray*}}
\newcommand{\enn}{\end{eqnarray*}}
\def\Ref#1{(\ref{#1})}

\ntitle{Non-Markov Statistical Effects of X-Ray Emission Intensity
of The Microquasar Grs 1915+105}

\nauthor{R.M. Yulmetyev$^{1,2}$, S.A. Demin$^{1,2}$, R.M.
Khusnutdinov$^{1,2}$, O.Yu. Panischev$^{1,2}$, P. H\"anggi$^3$}
\naddress{$^1$ Department of Physics, Kazan State University,
Kremlevskaya St. 18, Kazan, 420008, RUSSIA \\ $^2$Department of
Physics, Kazan State Pedagogical
University, Mezhlauk St. 1, Kazan, 420021, RUSSIA \\
{\small \rm E-mail:  rmy@theory.spu-kazan.ru;
sergey@theory.kazan-spu.ru}
\\ $^2$Department of Physics, University of Augsburg,
Universit\"atsstrasse 1, D-86135 Augsburg, GERMANY}

\ndata{ (Received 22 June 2005) }

\nabstract{In this paper we develop the new approach in time
series analysis with a variable time step and present the results
of quantitative and qualitative estimation of randomness and
regularity, and the study of non-Markovian effects of the X-ray
emission intensity of the microquasar GRS 1915+105. Our estimation
is based on the application of the theory of discrete
non-Markovian stochastic processes and gives a wide set of
quantitative characteristics and parameters of the studied system
with a variable time step. This set reflects the change of the
effects of randomness and regularity, the alternation of Markovian
and non-Markovian processes in the initial time signals. Initially
these characteristics were determined for the simple model systems
constructed by means of a molecular dynamics method. Further the
calculation of these parameters was executed for the X-ray
emission intensity of the microquasar GRS 1915+105. We have
developed a theoretical scheme for the detailed analysis of the
time series with a variable time step. The new theory allows us to
extract more detailed information about the X-ray emission of the
microquasar GRS 1915+105 from the experimental time series. The
received set of the parameters allows to estimate accurately
randomness and stochastic behavior, regularity and robustness,
non-Markovian effects of statistical memory in the intensity of
the X-ray emission. The analysis of the experimental data with a
constant and variable time step is carried out by means of phase
portraits, power spectra of the initial time and event correlation
function (TCF and ECF), and memory functions (MF) of the junior
order, the first three points of non-Markovity parameter.}

\nkeywords{discrete non-Markov processes, time series analysis,
randomness and non-Markovity, X-ray intensity of GRS 1915+105}

\nPACSnumbers{05.40.Ca; 89.75.-k; 95.75.Wx; 97.80.Jp}

\begin{document}

\DeclareGraphicsExtensions{.jpg,.pdf,.mps,.png} \firstpage{313}
\nlpage{330} \nvolume{9} \nnumber{4} \nyear{2006}
\def\nfpage{\thepage}
\thispagestyle{myheadings} \npcstitle

\section{Introduction}

The problems of randomness and stochastic behavior, regularity and
robustness have been in the focus of attention in the studies of
real complex systems of various nature over the past years. The
analysis of individual properties and characteristics of real
complex systems is impossible without registration and
quantitative estimation of various information on manifestation of
chaosity and randomness. By the change of the measure of
randomness or regularity it is possible to judge about the complex
dynamics of the system and its evolution. The discovery of the
phenomenon of chaos in dynamic systems has allowed to take a new
look at functioning of complex systems. Chaos is absence of order
and it characterizes randomness and unpredictability of changes in
the behavior of a system and impossibility to determine their
origin and  reasons.

While studying multiform manifestations of randomness, the authors
have defined chaos, and its the parameters describing chaotic and
regular stochastic processes in different ways. As a rule, the
search for the most accurate criteria for the estimation of
randomness or regularities in the dynamics of real objects is
carried out if we have the information on the behavior of the
well-known models of nonlinear dynamics such as, for example,
Lorentz model, Poisson model etc. Initially for the similar
systems one calculates the parameters determining their chaotic
dynamics. Further the given parameters are calculates for real
physical objects. On the basis of the comparative analysis of
these characteristics it is possible to come to certain
conclusions about the evolution and further dynamics of the
studied system. However the above mentioned models of nonlinear
dynamics do not carry sufficient information about the internal
properties of real objects. The models similar to real physical
objects allow to avoid these defects. The model description of a
gas, liquid and  solid by the methods of molecular dynamics
simulations can serve as an example.

In addition for the estimation of optimal qualitative and
quantitative parameters and characteristics of non-Markovian
effects and effects of randomness and regularity of X-ray emission
intensity dynamics of the microquasar GRS 1915+105 we use the
method of molecular dynamics. The galactic microquasar GRS
1915+105 was discovered as an X-ray transient in 1992
\cite{Castro-Tirado}, and has been observed to be extremely
luminous ever since. This binary system contains a 14 $M_{\odot}$
black hole \cite{Greiner_Nature} accreting from a late-type giant
of mass $0.8\pm 0.5$ $M_{\odot}$ \cite{Harlaftis} via Roche lobe
overflow. GRS 1915+105 is unique among accreting Galactic black
holes spending much of its time at super-Eddington luminosities
\cite{Done}. It is an extremely variable source, exhibiting
dramatic, aperiodic variability on a wide range of timescales,
from milliseconds to months \cite{Truss}. GRS 1915+105 is located
on the Galactic plane at a distance of $\sim 11 - 12$ $kpc$
\cite{Greiner_Nature, Kaiser} and suffers a large extinction of
25-30 mag in the visual band. The basic characteristics of the GRS
1915+105 binary system is the systemic velocity which is
$\gamma=-3\pm10$ $km/s$, the orbital period of the system is
$T_{orb}=33.5\pm1.5$ days \cite{Greiner_Astrophysics}.
Spectroscopic observations in the near-infrared H and K bands
identified absorption features from the atmosphere of the
companion (mass-donating star) in the GRS 1915+105 binary. The
detection of $^{12}CO$ and $^{13}CO$ band heads plus a few
metallic absorption lines suggested a $K-M$ spectral type and
luminosity class III (giant) \cite{Greiner_Nature}. Hard X-ray
studies in $20-100 keV$ band have shown erratic intensity
variations on time scales of days and months
\cite{Foster}-\cite{Paul}. In Ref. \cite{Greenhough_MNRAS} the
technique of differentiating and rescaling was applied to the GRS
1915+105 X-ray data. As a result the existence of a fundamental
time-scale for the system in the range of 12-17 days was found. In
Ref. \cite{Banibrata} it was concluded, that microquasar
GRS1915+105, as any other black hole system, may be chaotic in
nature. One of proofs of the similar behavior of black hole
systems is chaotic variability of the X-ray emission
\cite{Nayakshin}. For the estimation of a degree of randomness or
regularity of the X-ray emission of the microquasar GRS 1915+105
the correlation dimension of the system \cite{Banibrata} is often
used.

In this work we present a new method of research of randomness,
regularity, robustness and non-Markovian effects in the X-ray
emission intensity of the microquasar GRS 1915+105 on the basis of
the theory of discrete non-Markovian stochastic processes
\cite{Yulmetyev1}-\cite{Yulmetyev3}. The effects of non-Markovity
in real complex systems, natural \cite{Yulmetyev1, Yulmetyev2},
live \cite{Yulmetyev1}, \cite{Yulmetyev3}-\cite{Gafarov},
biological \cite{Goychuk}-\cite{Timmer} and physical
\cite{Gammaitoni, Mokshin} are of special interest for the
correlation analysis. This method allows to define the set of
characteristics and parameters, which contain detailed information
about non-Markovian effects and degrees of randomness of the X-ray
emission of the researched system. Initially we have calculated
the given characteristics and parameters for model systems by the
method of molecular dynamics: for a low density gas with greater
randomness and the stochastic behavior in particle motion; for a
high density gas; for a liquid near the triple point; for a solid
that corresponds to a greater regularity in movement of particles.
On the basis of the received results we have executed qualitative
and quantitative estimation of non-Markovian effects and of
stochastic and regular regimes in the initial signal with a
constant and variable time step for the X-ray emission intensity
of the microquasar GRS 1915+105.

\section{The theoretical framework of the statistical theory
of discrete non-Markovian stochastic processes}

In this section we present the basic concepts and definitions of
the theory of discrete non-Markovian stochastic processes
\cite{Yulmetyev1}-\cite{Yulmetyev3}, used  here for the analysis
of a time series with a constant and variable time step. The
theory is constructed on the discrete finite-difference
presentation of the Zwanzig-Mori kinetic equations
\cite{Zwanzig,Mori} well-known in statistical physics. The theory
of the discrete non-Markovian stochastic processes is widely
applied to the analysis of real systems of physical, biological,
live and social nature
\cite{Yulmetyev1}-\cite{Gafarov},\cite{Mokshin}. The dynamic,
kinetic and relaxation parameters and characteristics of this
theory, contain detailed information about individual properties
and qualities of the studied complex system.

\subsection{The basic concepts and definitions of the statistical theory
of discrete non-Markovian stochastic processes for the analysis of
time series with a constant time step}

As a rule, the time registration of any characteristics or
parameters of a complex system is carried out at discrete time
intervals of equal length. It allows to show the essential
features of fluctuations and irregularities and also the features
of various dynamic regimes in the initial time series. Some key
definitions and concepts of the statistical theory of discrete
non-Markovian stochastic processes for the analysis of the time
series with a constant time step
\cite{Yulmetyev1}-\cite{Yulmetyev3} are given below.

Let us present the time dynamics of the X-ray emission intensity
of the microquasar GRS 1915+105 as a discrete time series $x_j$ of
some variable X: \bn X=\{x(T),x(T+\tau),x(T+2\tau),\ldots,\\
\nonumber x(T+k\tau),\ldots,x(T+(N-1)\tau)\}. \label{Eq_1} \en
Here $T$ is the initial moment of the time the registration of the
X-ray emission intensity, $(N-1)\tau$ is the total time of
registration of the signal, $\tau$ is a discretization time step.
In the researched time series the discretization  time $\Delta
t=\tau=1$ day. The mean value of the variable $\langle X \rangle$,
fluctuation $\delta x_j$, and dispersion $\sigma^2$ can be
presented as follows: \bn
\langle X \rangle=\frac{1}{N}\sum_{j=0}^{N-1}x(T+j\tau), \\
\nonumber \delta x_j=x_j-\langle X \rangle, ~
\sigma^2=\frac{1}{N}\sum_{j=0}^{N-1}\delta x_j^2. \label{Eq_2}\en

For the description of the dynamic properties of the studied
complex system (the dynamics of correlations) it is convenient to
use the normalized time correlation function (TCF): \bn
a(t)=\frac{1}{(N-m)\sigma^{2}}\sum_{j=0}^{N-1-m}\delta
x(T+j\tau)\\ \label{Eq_3} \nonumber \times \delta x(T+(j+m)\tau),
\en where $t=m\tau$.
 Here $\delta x_j, \delta x_{j+m}$ are fluctuations
of the variable $X$ at $j, j+m$ step, correspondingly,
$\sigma^{2}$ is an absolute dispersion of the variable $X$. TCF in
Eq. (3) satisfies the requirements of normalization and
attenuation of correlations: \be \lim_{t\rightarrow 0}a(t)=1,~
\lim_{t\rightarrow \infty}a(t)=0. \label{Eq_4} \ee It is necessary
to note, that for a certain class of complex systems the
requirement of correlations attenuation is not always possible.

By the use of the Zwanzig-Mori projection operators technique
\cite{Zwanzig,Mori} it is possible to receive an interconnected
chain of finite-difference equations of a non-Markovian type
\cite{Yulmetyev1}-\cite{Yulmetyev3} for the initial TCF $a(t)$ and
the memory function $ M_{i}(t) $ ($ i=1,2,...,n $): \bn
\frac{\Delta a(t)}{\Delta
t}=-\tau\Lambda_{1}\sum_{j=0}^{m-1}M_{1}(j\tau)a(t-j\tau)\nonumber
\\+\lambda_{1}a(t), \nonumber \\ \frac{\Delta M_{1}(t)}{\Delta
t}=-\tau\Lambda_{2}\sum_{j=0}^{m-1}M_{2}(j\tau)M_{1}(t-j\tau)\nonumber
\\+\lambda_{2}M_{1}(t),
\nonumber \\ \ldots, \nonumber \\ \frac{\Delta M_{n-1}(t)}{\Delta
t}=-\tau\Lambda_{n}\sum_{j=0}^{m-1}M_{n}(j\tau)M_{n-1}(t-j\tau)
\nonumber\\+\lambda_{n}M_{n-1}(t),\label{Eq_5}
 \en where $\Lambda_{i}$ are relaxation parameters,
and parameters $\lambda_{i}$ form the eigenvalue spectrum of
Liouville's quasioperator  $\hat{L}$: \be
\lambda_{n}=i\frac{<\textbf{W}_{n-1}\hat{L}\textbf{W}_{n-1}>}{<|\textbf{W}_{n-1}|^{2}>},
\label{Eq_6} \ee \be
\Lambda_{n}=i\frac{<\textbf{W}_{n-1}\hat{L}\textbf{W}_{n}>}{<|\textbf{W}_{n-1}|^{2}>}.
\label{Eq_7} \ee Orthogonal dynamic variables $\textbf{W}_n$ are
received with the help of the Gram-Schmidt orthogonalization
procedure:

\be \langle \textbf{W}_n \textbf{W}_m \rangle=\delta_{n,m} \langle
|\textbf{W}_n|^2 \rangle, \nonumber \ee where $\delta_{n,m}$ is
Kronecker's symbol. To compare the relaxation time scales for the
initial TCF $a (t) $ and the $i$th order memory functions $M _ {i}
(t)$ we use the non-Markovity statistical parameter $ \varepsilon
$. Initially the given parameter was used for the analysis of the
irreversible phenomena in a condensed matter
\cite{Shurygin}-\cite{Yulmetyev5}. According to \cite
{Yulmetyev1}-\cite{Yulmetyev3} we introduce the relaxation times
of the initial TCF and the $n$th order memory functions: \bn
\tau_a = \triangle t \sum_{j=0}^{N-1}a(t_j), \nonumber \\
\tau_{M_1} = \triangle t \sum_{j=0}^{N-1}M_1(t_j),\nonumber\\
\ldots, \nonumber\\\tau_{M_n} = \triangle t
\sum_{j=0}^{N-1}M_n(t_j). \label{Eq_8} \en Then the spectrum of
the non-Markovity parameter is defined as a set of dimensionless
quantities \cite {Shurygin}: \bn
\{\varepsilon_i\}=\{\varepsilon_1,\varepsilon_2,...,\varepsilon_{n-1}\},\nonumber
\\ \varepsilon_1=\tau_a/\tau_{M_1},
\varepsilon_2=\tau_{M_1}/\tau_{M_2},  \ldots,\nonumber \\
\varepsilon_{n}=\tau_{M_{n-1}}/\tau_{M_{n}}. \label{Eq_9} \en
Thus, the value $ \varepsilon_n $ characterizes the comparison of
relaxation times of the memory functions $M _ {n-1} $ and $M_n$.
The non-Markovity parameter allows to divide all relaxation
processes into Markov, quasi-Markov and non-Markov processes. The
spectrum of the non-Markovity parameter defines the stochastic
peculiarities of TCF.

In work \cite {Yulmetyev1} the concept of the non-Markovity
generalized parameter for frequency - dependent case was
introduced: \bn \varepsilon_i (\nu) = \left\{ \frac {\mu _ {i-1}
(\nu)} { \mu_i (\nu)} \right\}^{\frac {1} {2}}. \label{Eq_10} \en
where $\mu_i (\nu)$ is a power spectrum of the $i$th order memory
function: \bn \mu_0(\nu)=|\triangle t
\sum_{j=0}^{N-1}a(t_j)cos2\pi\nu t_j|^2,\nonumber \\ \ldots,
\nonumber \\ \mu_i(\nu)=|\triangle t
\sum_{j=0}^{N-1}M_i(t_j)cos2\pi\nu t_j|^2. \nonumber \en

The above-mentioned equations (3)-(5) present the case of
Zwanzig-Mori's statistical theory \cite {Zwanzig, Mori} for the
discrete statistical complex systems. The statistical theory of
discrete non-Markovian stochastic processes allows to reveal
Markov and non-Markov effects, the effects of statistical memory,
effects of a dynamic alternation of stochastic and regular regimes
in the initial time series for the X-ray emission intensity of the
microquasar GRS 1915+105.

\subsection{A new approach to the analysis of discrete
time series with a variable time step}

In many real complex systems the registration of the initial time
signal by different reasons is carried out at time intervals of
different length. To such systems we can refer the objects of an
astrophysical and seismological nature, some biological and social
systems, economic and ecological objects \cite{Tirnakli, Abe1,
Abe2}.

In the given work we offer a new approach to the description of
discrete non-Markovian stochastic processes with a time step of
variable length. Such presentation of the initial time signal
allows to find the dynamic development of the system, connected
with a not real time scale but with its consistent presentation.
The basic idea of this method consists in fixing individual events
as a sequence of dynamic values. It allows to consider the
dynamics of the system as a sequence of individual events. As an
example, here we present the analysis of the time registration
with a variable time step of the X-ray emission intensity of the
microquasar GRS 1915+105.

\subsubsection{\emph{2.2.1 The basic concepts and definitions of the
theory of discrete non-Markovian stochastic processes with a
variable time step}}

Let us consider the chaotic dynamics of the X-ray emission
intensity as a sequence of events which are ``non-uniform'' on a
time scale:
\begin {eqnarray}
E=\{ \xi_{1},\xi_{2},\xi_{3},\ldots,\xi_{k},\ldots,\xi_{N}\},
\label{Eq_11}
\end {eqnarray}
where the intervals of time $\Delta t_{ij}=t_{i}-t_{j},j=i+1,
j=i-1 $, are unequal. Here $\xi_{i}$ presents the event at the
moment $t_{i}$ which follows after the event $\xi_{i-1}$,
$i=1,..., N $ is the number of the event.

The mean value $\langle E \rangle$, fluctuation $\delta \xi_{i}$,
absolute dispersion $\sigma^{2}$ for the set of the random
variable $E$ are defined as follows: \bn \langle E
\rangle=\frac{1}{N}\sum_{i=1}^{N} \xi_{i}, ~ \delta
\xi_{i}=\xi_{i}-\langle E \rangle,
 \nonumber\\
\sigma^{2}=\frac{1}{N}\sum_{i=1}^{N}\delta
\xi_{i}^{2}=\frac{1}{N}\sum_{i=1}^{N}\{ \xi_{i}-\langle E
\rangle\}^{2}.
 \label{Eq_12}
\en According to \cite{Tirnakli}-\cite{Abe2} we shall define the
correlation dependence in a sequence of events (11) as follows:
\bn a(n)=\frac{1}{(N-m)\sigma^{2}}\sum_{i=1}^{N-m}\delta \xi_{i}
\delta \xi_{i+m}. \label{Eq_13} \en The function introduced in the
similar way $a(t)$ presents the event (not time!) correlation
function (ECF). The general requirements suggest, that ECF should
have the properties of normalization and attenuation of
correlations: \be \lim_{n\rightarrow 1}a(n)=1,~ \lim_{n\rightarrow
\infty}a(n)=0. \label{Eq_14} \ee For the description of a discrete
sequence of events we shall use Liouville's finite-difference
equation of movement: \be \frac{\Delta \xi_{i}(n)}{\Delta
n}=i\widehat{L}(n,1)\xi_{i}(n),
 \label{Eq_15}
\ee
where $\xi_{i}(n+1)=U(n+1,n)\xi_{i}(n)$, $U(n+1,n)$ is an
evolution operator.

Thus, the left hand side of the Eq. (15) can be submitted as
follows: \bn \frac{\Delta \xi_{i}}{\Delta
n}=U(n+1,n)\xi_{i}(n)-\xi_{i}(n)=\nonumber\\\{U(n+1,n)-1
\}\xi_{i}(n). \label{Eq_16} \en Let us present the set of values
of the dynamic variable $\delta \xi_{i}$ as a $k$-component vector
of the system's state:
\\a) the vector of the initial state:
\begin{subequations}
\be \textbf{A}_{k}^{1}=\{ \delta \xi_{1},\delta \xi_{2},\delta
\xi_{3},...,\delta \xi_{k} \}, \label{Eq_17a} \ee b) the vector of
the final state: \be \textbf{A}_{m+k}^{m}=\{ \delta \xi_{m},\delta
\xi_{m+1},\delta \xi_{m+2},...,\delta \xi_{m+k} \}, \label{Eq_17b}
\ee
\end{subequations}
 where $1\leq k \leq N$.

Using the standard expression for the scalar product of vectors
and relations (13), (17a) and (17b), we receive the ``event''
correlation function (ECF) for a stationary sequence of events
\Ref{Eq_11} in the following way: \be a(n)=\frac{\langle
\textbf{A}_{k}^{1}(1)\textbf{A}_{k+m}^{m}(n)\rangle}{\langle
|\textbf{A}_{k}^{1}(1)|^{2}\rangle}. \label{Eq_18}\ee

\subsubsection{\emph{2.2.2 Kinetic equations for discrete non-Markov processes
with variable time step}}

Let us write down the equation of motion (15) for the vector
state: \be \frac{\Delta \textbf{A}_{m+k}^{m}(n)}{\Delta
n}=i\widehat{L}(n,1)\textbf{A}_{m+k}^{m}(n). \label{Eq_19}\ee By
the use of the projection operator technique we can split
Euclidean vector's space of state $\textbf{A}(k)$ into two
mutually-orthogonal subspaces: \bn
\textbf{A}(k)=\textbf{A}^{'}(k)+\textbf{A}^{''}(k), \nonumber\\
\textbf{A}^{'}(k)=\Pi \textbf{A}(k),~ \textbf{A}^{''}(k)=P
\textbf{A}(k). \label{Eq_20}\en Here projector operators $\Pi$ and
$P$ have the following properties: \bn
\Pi=\frac{|\textbf{A}_{k}^{1}(1)\rangle \langle
\textbf{A}_{k}^{1}(1)|}{\langle
|\textbf{A}_{k}^{1}(1)|^{2}\rangle}, \nonumber\\ P=1-\Pi, ~
\Pi=\Pi^{2}, \nonumber\\ P^{2}=P, ~ \Pi P=P\Pi=0. \label{Eq_21}\en
It allows to split Liouville's equation (15) into two appropriate
equations in two orthogonal subspaces:
\begin{subequations}
\be \frac{\Delta \textbf{A}^{'}(n)}{\Delta
n}=i\widehat{L}_{11}\textbf{A}^{'}(n)+i\widehat{L}_{12}\textbf{A}^{''}(n),
\label{Eq_22a} \ee \be \frac{\Delta \textbf{A}^{''}(n)}{\Delta
n}=i\widehat{L}_{21}\textbf{A}^{'}(n)+i\widehat{L}_{22}\textbf{A}^{''}(n).
\label{Eq_22b} \ee
\end{subequations}
Here $\widehat{L}_{ij}=\Pi_{i}\widehat{L}\Pi_{j}$ presents the
matrix elements of Liouville's quasioperator
$\widehat{L}=\widehat{L}_{11}+\widehat{L}_{12}+\widehat{L}_{21}+\widehat{L}_{22}$.
Solving the Eq. (22b) and substituting into the Eq. (22a), we
shall receive: \bn \frac{\Delta \textbf{A}^{'}(n+m)}{\Delta
n}=i\widehat{L}_{11}\textbf{A}^{'}(n+m)\nonumber\\+i\widehat{L}_{12}\{
1+i\Delta n\widehat{L}_{22}\}^{m}\textbf{A}^{''}(n) \nonumber\\-
\widehat{L}_{12}\sum_{j=1}^{m}\{1+i\Delta
n\widehat{L}_{22}\}^{j}\Delta
n\widehat{L}_{21}\textbf{A}^{'}(n+[m-j]).
 \label{Eq_23}
\en Acting on the Eq. (23) by the operator of $\langle
\textbf{A}(1)|/\langle |\textbf{A}(1)|^{2}\rangle$ and taking into
account an idempotentity property of projection operators, we can
receive the following finite-difference equation for the initial
event correlation function: \be \frac{\Delta a(n)}{\Delta
n}=i\lambda_{1}a(n)-\Delta
n\Lambda_{1}\sum_{j=1}^{m}M_{1}(j)a(n-j).
 \label{Eq_24}
\end {equation}
Assuming that $ \Delta n=1 $, it is possible to formally solve
this equation:
\begin {equation}
a(n+1)=\{i\lambda_{1}+1\}a(n)-\Lambda_{1}\sum_{j=1}^{m}M_{1}(j)a(n-j).
\nonumber
\end {equation}

Here $\lambda_{1}$ is an eigenvalue Liouville's quasioperator.
Relaxation parameter $\Lambda_{1}$ and memory function $M_{1}(j)$
are defined as follows: \be \lambda_{1}=\frac{\langle
\textbf{A}_{k}^{1}(1)\widehat{L}\textbf{A}_{k}^{1}(1)\rangle}{\langle
|\textbf{A}_{k}^{1}(1)|^{2}\rangle}, \nonumber \ee \be
\Lambda_{1}= \frac{\langle
\textbf{A}_{k}^{1}(1)\widehat{L}_{12}\widehat{L}_{21}\textbf{A}_{k}^{1}(1)\rangle}{\langle
|\textbf{A}_{k}^{1}(1)|^{2}\rangle}, \nonumber\ee \be
M_{1}(j)=\frac{\langle \textbf{A}_{k}^{1}(1)\widehat{L}_{12}\{
1+i\Delta
n\widehat{L}_{22}\}^{j}\widehat{L}_{21}\textbf{A}_{k}^{1}(1)\rangle}{\langle
\textbf{A}_{k}^{1}(1)\widehat{L}_{12}\widehat{L}_{21}\textbf{A}_{k}^{1}(1)\rangle}.
\nonumber \ee Equation (24) contains function $M_{1}(j)$, for
which it is possible to repeat the procedure submitted above and
receive appropriate finite-difference equations of a non-Markovian
type for memory functions of senior orders $n>1 $. To simplify the
given procedure and generalize the received results one can use
Gram-Schmidt orthogonalization procedure \cite{Reed}:

\be \langle \textbf{W}_{s}
\textbf{W}_{p}\rangle=\delta_{s,p}\langle
|\textbf{W}_{s}|^{2}\rangle. \nonumber\ee This operation allows to
receive a new vector of state $\textbf{W}_{s}$, contained in the
memory function $M_{s}(j)$: \bn
\textbf{W}_{0}=\textbf{A}_{k}^{1},~ \textbf{W}_{1}=\{
i\widehat{L}-\lambda_{1}\}\textbf{W}_{0},\nonumber\\
\textbf{W}_{2}=\{
i\widehat{L}-\lambda_{2}\}\textbf{W}_{1}-\Lambda_{1}\textbf{W}_{0},
\ldots \label{Eq_25}\en For new orthogonal dynamic variables
$\textbf{W}_{s}$ we receive an interconnected chain of
finite-difference equations of a non-Markovian type for the $s$th
order normalized correlation functions: \bn
 \frac{\Delta M_{1}(n)}{\Delta
n}=-\Delta n\Lambda_{2}\sum_{j=1}^{m}M_{2}(j)M_{1}(n-j)\nonumber
\\+i\lambda_{2}M_{1}(n),
\nonumber \\
 \ldots,  \nonumber \\
 \frac{\Delta M_{s}(n)}{\Delta
n}=-\Delta
n\Lambda_{s+1}\sum_{j=1}^{m}M_{s+1}(j)M_{s}(n-j)\nonumber \\
+i\lambda_{s+1}M_{s}(n), \label{Eq_26} \en where \bn
M_{1}(n)=\frac{\langle
\textbf{W}_{1}(1)\textbf{W}_{1}(n)\rangle}{\langle
|\textbf{W}_{1}(1)|^{2}\rangle}, \nonumber \\ \ldots,
\nonumber\\M_{s}(n)=\frac{\langle
\textbf{W}_{s}(1)\textbf{W}_{s}(n)\rangle}{\langle
|\textbf{W}_{s}(1)|^{2}\rangle}, \nonumber \en

\bn \lambda_{s+1}=\frac{\langle
\textbf{W}_{s}[i\widehat{L}\textbf{W}_{s}] \rangle}{\langle
|\textbf{W}_{s}|^{2}\rangle},  \nonumber\\ \ldots, \nonumber\\
\Lambda_{s+1}=\frac{\langle
\textbf{W}_{s}[i\widehat{L}\textbf{W}_{s+1}] \rangle}{\langle
|\textbf{W}_{s}|^{2}\rangle}. \label{Eq_27}\en

The frequency - dependence of statistical spectrum of the
non-Markovity parameter for the case of time series with a
variable time step will be defined as follows: \be
\varepsilon_{i}(\nu)=\left \{\frac{\mu_{i-1}(\nu)}{\mu_{i}(\nu)}
\right \}^{\frac{1}{2}}, \label{Eq_28}\ee where $\mu_{i}(\nu)$ is
a power spectrum for the $i$th order correlation function: \bn
\mu_{1}(\nu)=\left \{ \Delta n \sum_{n=1}^{N}M_{1}(n)\cos(2\pi
n\nu)\right \}^{2},\nonumber\\\ldots,\nonumber\\
\mu_{i}(\nu)=\left \{\Delta n \sum_{n=1}^{N}M_{i}(n)\cos(2\pi
n\nu)\right \}^{2}. \label{Eq_29}\en

\section{The experimental data and the details of computer simulations}

We analyze here two types of the experimental data of the X-ray
emission intensity of the microquasar GRS 1915+105 \cite{data}.
The first set of the data presents a one-day averaged time series
of the X-ray emission intensity in the period from February, 1,
1996 to September, 1, 2004 (a step discretization $\tau=1$ day).
For the analysis of the given time series we use the statistical
theory of discrete non-Markov processes with a constant time step
(see Sect. 2.1).

For the analysis of the non-equidistant time series (the 2nd type
of data) we use the statistical theory of discrete non-Markovian
stochastic processes with a variable time step (see Sect. 2.2). As
experimental data \cite{data} we use the time series of the X-ray
emission intensity of the microquasar GRS 1915+105 with a variable
time step for the period from February, 1, 1996 to September, 1,
2004. To study the stochastic properties of the X-ray emission we
have carried out an additional study of four model systems using
the method of molecular dynamics. We have studied a model system
(a gas, a liquid, and a solid) consisting of 2048 particles of
argon molecules. The particles interacted by Lennard-Jones
potential
$V(r)=4\varepsilon\{({\sigma}/{r})^{12}-({\sigma}/{r})^{6}\}$ with
parameters $\varepsilon/{k_{B}}=120 K$ and $\sigma=3.405
\textrm{\AA}$ \cite{Rahman}. Here $\varepsilon$ is a well depth,
$\sigma$ is a interatomic distance. The simulation was carried out
at constant temperature $T^{*}=0.6$ and at various densities
$n^{*}=0.1$, $n^{*}=0.5$, $n^{*}=1.0$ and $n^{*}=1.5$. We made use
of the ``velocity Verlet'' algorithm to integrate the equations of
motion \cite{Verlet} with a time step $10^{-14}$ sec.

\section{Randomness, regularity and non-Markov effects
applied to the dynamic analysis of simple model systems and the
X-ray emission intensity of the microquasar GRS 1915+105}

In this section we present a new method of quantitative estimation
of randomness, regularity and non-Markov effects of a time series.
Preliminary the calculation of the degree of manifestation of
non-Markov effects and randomness or regularities in the dynamic
movement of a particle in the given cell will be carried out on
simple model systems: a low density gas, a dense gas, a liquid
near triple point, a solid. The level of randomness and non-Markov
effects is established for each model system and is carried out by
means of a set of various characteristics. The set of data of
characteristics and quantitative parameters contains reliable
information about the degree of randomness or regularity and
non-Markov effects in the researched model system. Then on the
basis of the calculation of these characteristics we shall proceed
to the analysis of a real process: the event variability of the
X-ray emission intensity of the microquasar GRS 1915+105.

\subsection{Randomness and non-Markov effects in simple model systems}

The study of the dynamic features of behavior of real complex
systems of a different nature in cardiology, neurophysiology,
epidemiology, biophysics, seismology shows the existence of close
connection between the first point of the non-Markovity parameter
and a quantitative measure of randomness or regularity of the
measured signal. To establish this connection we present the
results of the study of several simple physical models. The
purpose of the similar study consists in detecting typical
features of behavior of the non-Markovity parameter for model
systems with a different degree of randomness or regularity. The
given models were constructed by the method of molecular dynamics.
As an example we have considered four Lennard-Jones model systems:
a low density gas ($T^{*}=0.6$, $n^{*}=0.1$ in reduced units with
parameters, where $\sigma=3.405 \textrm{\AA}$ is an interatomic
distance, $\varepsilon/k_B=120 K$ is the well depth;
$T=T^*\varepsilon/k_B, n=n^*/\sigma^3$); a dense gas ($T^{*}=0.6$,
$n^{*}=0.5$); a liquid near the triple point ($T^{*}=0.6$,
$n^{*}=1.0$) and a solid ($T^{*}=0.6$, $n^{*}=1.5$).

\begin{figure}[ht!]
     \leavevmode
\centering
\includegraphics[width=3.3in, height=2.5in, angle=0]{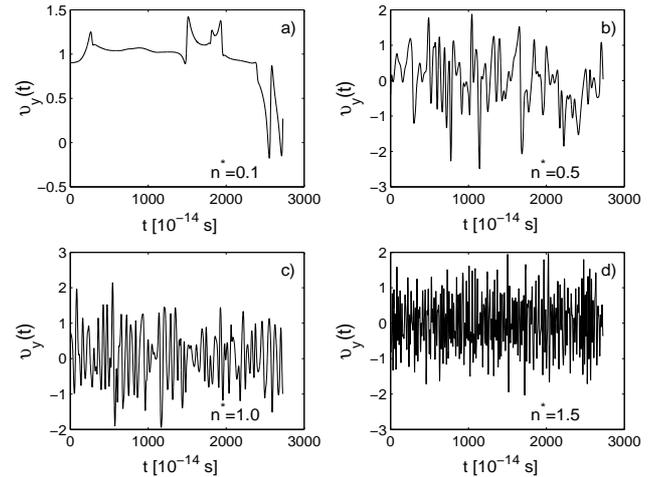}
\caption{The time dependence of the $y$--component of particle
velocity: a) for a low density gas ($T^{*}=0.6$, $n^{*}=0.1$, in
reduced units); b) for a dense gas ($T^{*}=0.6$, $n^{*}=0.5$); c)
for a liquid near the triple point ($T^{*}=0.6$, $n^{*}=1.0$); d)
for a solid ($T^{*}=0.6$, $n^{*}=1.5$). The weakest correlations
correspond to the $y$--component of the particle velocity dynamics
for a low density gas. This dynamics characterizes the most
chaotic behavior of the particle motion in the studied models.
With the increase of density of the model system the correlation
in the behavior of particles become more intensive. A model of a
solid corresponds to the greatest probability of interaction
between particles and the strongest velocity fluctuations.}
\end{figure}

The time dependencies of the $y$-component velocity for one
particle in the studied cell is submitted in Fig. 1. On the basis
of the comparative analysis we can reveal a clear distinction
between the behavior of particles in each model. For the case of a
low density gas (see Fig. 1а) a weak correlation between the
particle velocity and time is observed. It is connected with great
(in comparison with the size of particles) distances between
particles and the collisionless regime of their behavior. Within a
long-time limit weak correlations caused by interaction of two
(three) particles are detected. In the given model the most
chaotic behavior of a particle in all the studied systems can be
observed. Thus, the $y$-component of a particle velocity of a
dense gas (see Fig. 1b) differs in strong correlations with time.
The interval of fluctuation scattering is relatively fixed. The
model shows ``moderate chaotization'' in the behavior of
particles. The following model corresponds to a liquid near the
triple point (see Fig. 1c) is characterized by the state of
``moderate regularity'' in the motion of particles. In comparison
with the previous models the correlation between the motion of
 molecules amplifies noticeably. The amplification of correlations
is connected to the increase of the density of the system,
accordingly, to greater intensity of interaction of particles. The
amplitude of fluctuations is found within a certain interval of
values. The obvious regularity in the motion of real molecules
corresponds to the model of a solid state (see Fig. 1d). The
diagram shows appreciable symmetry of fluctuations regarding the
value characterizing the condition of equilibrium in the given
model. A high degree of regularity is defined by a high degree of
correlation of the states of particles.

\begin{figure}[ht!]
     \leavevmode
\centering
\includegraphics[width=3.3in, height=2.5in, angle=0]{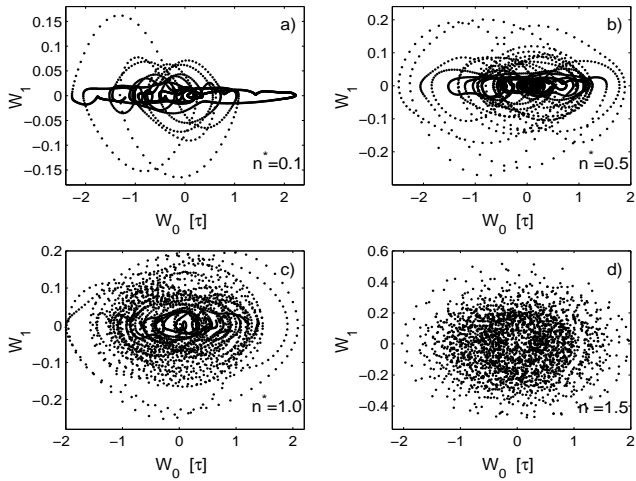}
\caption{The phase portraits on a plane projection of the two
first orthogonal dynamic variables ($W_0, W_1$) for the considered
model systems: a) for a low density gas; b) for a dense gas; c)
for a liquid near the triple point; d) for a solid. The structure
of phase clouds indicates a level of randomness or regularity of
the studied objects. Higher symmetry and regular concentration of
phase points concerning the center of coordinates are
characteristic for objects with a high degree of regularity.
Decreasing of concentration of the phase points near center of
coordinates is connected with the increase of the level of
randomness in the system behavior.}
\end{figure}

The phase portraits on plane projections of the two first
orthogonal dynamic variables ($W_{0}, W_{1}$) are submitted in
Fig. 2. Here various degrees of randomness and regularity in the
movement of the particles in the system is observed. In all the
phase clouds various symmetry is revealed. In case of a low
density gas the individual points of the phase portrait generate
precise closed structures with a complex chaotic form. In the next
models the phase points concentrate near to the center of the
coordinate system. The density of the phase cloud grows with the
growth of regularity of the model. Concentrated orbits, located
around the central nucleus, dissipate. The phase portraits of the
solid model are characterized by the greatest concentration of the
phase points near the origin of coordinates. It is necessary to
note that with the transition of the system from a) to c) the
fluctuation scale of variable $W _ {2} $ increases. The degree of
changes of the orthogonal dynamic variables noticeably differs
owing to the difference between the degree of correlations of
particles.
\begin{figure}[ht!]
     \leavevmode
\centering
\includegraphics[width=3.3in, height=2.5in, angle=0]{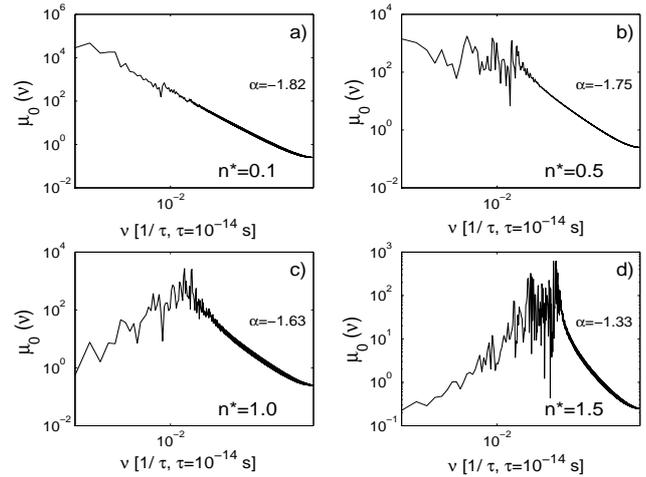}
\caption{The power spectrum of the initial time correlation
function $\mu_0(\nu)$ for four model systems. The parameter
$\alpha$ was calculated for each power spectrum. By estimating
values $\alpha$ for each model system it is possible to reveal its
connection with a level of randomness or regularity in the
behavior of the system.}
\end{figure}

In Fig. 3 the power spectrum of the initial time correlation
function $ \mu _ {0} (\nu) $ for four model systems is submitted.
The fractal parameter $\alpha$ of the power spectrum $\mu_{0}(\nu)
\sim \nu^{\alpha}$ (where $\alpha<0$ ) was calculated on all
frequency scale for all model systems. For a low density gas this
parameter amounts to $\alpha=-1.82$; for a gas $\alpha=-1.75$. For
a liquid it is equal to $\alpha=-1.63$; for a solid
$\alpha=-1.33$. Thus, between the fractal parameter of the power
spectrum $\mu _ {0}(\nu)$ and the level of manifestation of a
randomness and stochastic behavior (the parameter $\alpha$ grows
with the increase of a regularity degree) there is appreciable
interrelation.

\begin{figure}[ht!]
     \leavevmode
\centering
\includegraphics[width=3.3in, height=2.5in, angle=0]{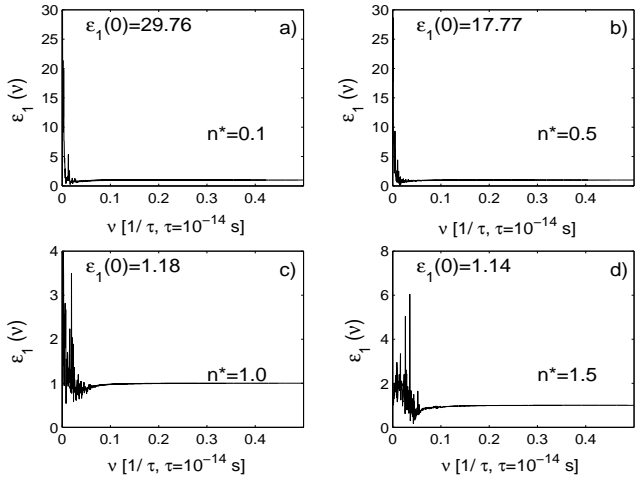}
\caption{The frequency dependence of the first point of the
non-Markovity parameter $\varepsilon_1(\nu)$ for the considered
model systems. The analysis of various physical and natural
systems specifies a special value of the parameter $\varepsilon_1
(0)$ in the estimation of level of randomness or regularity. The
parameter $\varepsilon_1 (0)$ make it possible to judge about the
level of randomness or regularity in the system. The maximal
values of parameter $\varepsilon_1(0)$ ($\sim 10^1 \div 10^2)$ are
characteristic for chaotic and fully randomized processes.
Decreasing of the parameter $\varepsilon_1(0)$ up to values $\sim
10^0$ reflects the increase of level of regularity. It is possible
to come to the similar conclusions by estimating the values of
this parameter for the submitted models.}
\end{figure}

In Fig. 4 the frequency dependence of the first point of the
non-Markovity parameter (further simply the non-Markovity
parameter), calculated by Eq. (10) is submitted.

While analyzing the majority of natural \cite{Yulmetyev1,
Yulmetyev2} and physical \cite{Mokshin} systems earlier we found
the following feature in the behavior of the first point of the
non-Markovity parameter. The value $\varepsilon_{1}(0) \sim 10^{1}
\div 10^{2}$ corresponds to the dynamic states of the systems with
the greatest level of randomness and stochastic behavior. For the
similar states pronounced Markov effects (manifestation of
instantaneous or short-range statistical memory effects) are
characteristic. With the increase of regularity and robustness in
the system the numerical value of the first point of the
non-Markovity parameter on zero frequency decreases up to a unit
$\varepsilon_{1}(0)\sim 10^{0}$. Non-Markov processes with
amplifying effects of long-range statistical memory correspond to
such states.

The analysis of the model systems lead to the similar results. The
parameter $\varepsilon _ {1}(0)$ for the model of a low density
gas (see Fig. 4а) constitutes 29.76. This is a maximal value of
the non-Markovity parameter characterizing a low density gas i.e.
the model with the greatest level of randomness and strong
Markovian effects. The value of this parameter
$\varepsilon_{1}(0)$ for a dense gas equal to 17.77, for a liquid
- 1.18, for a solid 1.14. With the increase of the density of the
system the numerical value of the non-Markovity parameter on zero
frequency decreases to a unit (with the increase of the density of
the system the non-Markovian effects increase). This testifies to
the possibility to use the non-Markovity parameter as an
informational measure of manifestation of randomness and
regularity effects.

Thus, the submitted model systems allow to define the set of
qualitative and quantitative characteristics for the analysis of a
degree of randomness and non-Markov effects in real complex
systems. These characteristics carry detailed information about
randomness or regularity in the researched system. To these
characteristics we can refer: the time correlation in the initial
time series, the shape of the phase clouds ($W_{1}=f(W_{0})$), the
fractal parameter of the power spectrum  $\mu_{0}(\nu)$ and the
first point of the non-Markovity parameter on zero frequency
$\varepsilon_{1}(0)$. The most authentic and informative parameter
of the level of randomness and the manifestation of non-Markov
effects is the statistical non-Markovity parameter.

\subsection{The analysis of the X-ray emission intensity of the
microquasar GRS 1915+105 for the series with a constant time step}

The time series of the X-ray emission intensity of the microquasar
GRS 1915+105 \cite {data} during the period from February, 1, to
September, 1, 2004 (the time discretization is equal to one day)
is submitted in Fig. 5. Here we find the presence of quasiperiodic
structures which are connected to a relative regularity of the
signal within certain time intervals. It the end of the time
series the quasiperiodic structures get most noisy therefore their
form is destroyed.

\begin{figure}[ht!]
     \leavevmode
\centering
\includegraphics[width=3.3in, height=2.5in, angle=0]{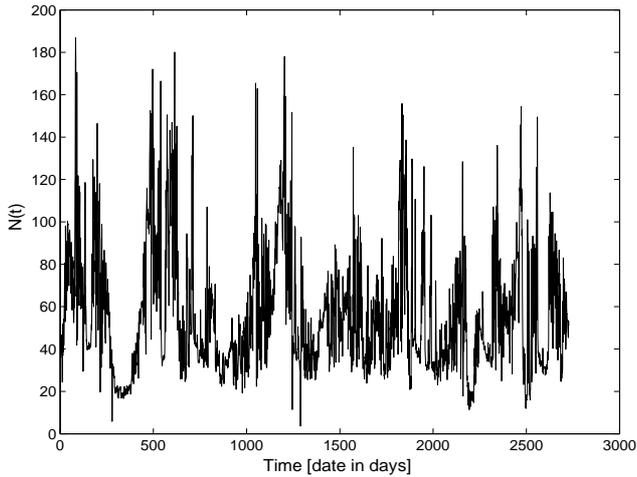}
\caption{The time dependence of the X-ray emission intensity of
the microquasar GRS 1915+105 with constant time step (step
discretization $\tau=1$ day). It is possible to discover
quasiperiodic structures in the time dependence. The given
structures bring relative regularity in the X-ray emission
intensity.}
\end{figure}

The phase clouds on the plane projections for various combinations
of dynamical orthogonal variables $W_i$, $W_j$ (where $i, j=0..3$)
of the X-ray intensity of the microquasar GRS 1915+105 are shown
in Fig. 6. It is possible to judge the character of the X-ray
emission intensity by the shape of the phase clouds. The phase
clouds are of an asymmetric type and consist of a centralized
nucleus with a high concentration of phase points and several
points surrounding the nucleus.

\begin{figure}[ht!]
     \leavevmode
\centering
\includegraphics[width=3.3in, height=2.5in, angle=0]{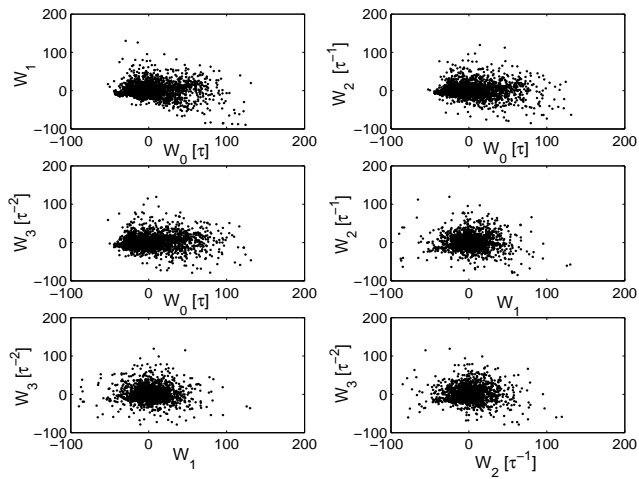}
\caption{The phase portraits on the plane projections of the four
first dynamic orthogonal variables of the X-ray emission intensity
of the microquasar GRS 1915+105 (for time series presented in Fig.
5). The phase clouds consist of the centralized nucleus with a
high concentration of phase points and individual points scattered
on the perimeter.}
\end{figure}

\begin{figure}[ht!]
     \leavevmode
\centering
\includegraphics[width=3.3in, height=2.5in, angle=0]{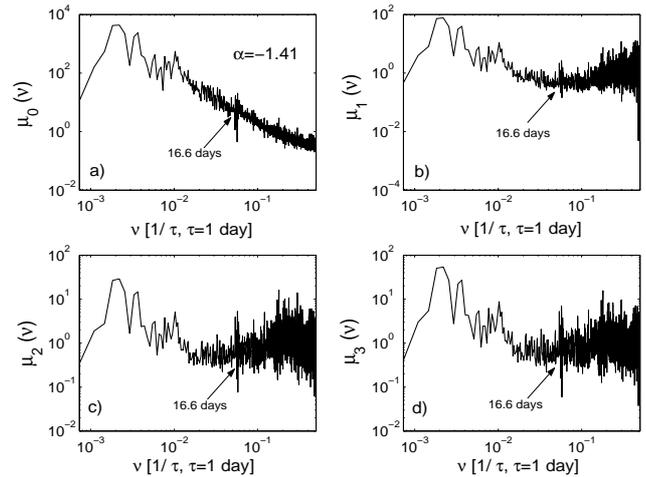}
\caption{The power spectrum of the initial TCF $\mu_0(\nu)$ (a)
and three memory functions of the junior order (b, c, d) for the
X-ray emission intensity of the microquasar GRS 1915+105 (for time
series presented in Fig. 5). The fractal parameter of the power
spectrum $\mu_0(\nu)$ is equal to $\alpha=-1.41$. This value is an
intermediate value between the similar parameters for the models
of a liquid ($\alpha=-1.63$) and a solid ($\alpha=-1.33$).}
\end{figure}

The power spectrum of the initial TCF $\mu_{0}(\nu)$ (Fig. 7a) and
three memory functions of the junior order $\mu_{i}(\nu)$ (where
$i=1,2,3$) (Figs. 7b, c, d) for the intensity of the X-ray
emission are depicted in Fig. 7. All the figures are submitted on
a double logarithmic scale. The fractal parameter of the power
spectrum $\mu_{0}(\nu)$ is equal to $\alpha=-1.41$. This value
corresponds to the intermediate quantity between the similar
parameters for the model of a liquid $\alpha=-1.63$ and a solid
$\alpha=-1.33$. In the region of frequencies $\nu$=$
5.7\times10^{-2} \div7.2\times10^{-2} f.u. (1 f.u. = 1/\tau$ where
$\tau$ is a discretization time step) a series of dynamic peaks is
detected in the power spectra. This frequency range corresponds to
a time interval of $\tau=13.8 \div 17.4$ days. The maximal peak
corresponds to frequency $\nu=6\times 10^{-2} f.u., $ of
$\tau=16.6$ days.

\begin{figure}[ht!]
     \leavevmode
\centering
\includegraphics[width=3.3in, height=1.6in, angle=0]{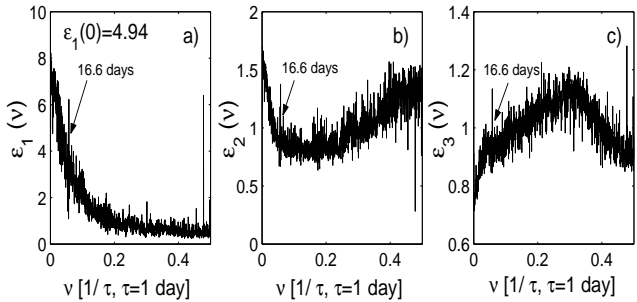}
\caption{The frequency dependence of the first three points of the
non-Markovity parameter $\varepsilon_i (\nu)$ ($i=1,2,3$) for the
X-ray emission intensity of the microquasar GRS 1915+105 (for time
series presented in Fig. 5). The parameter $\varepsilon_1 (0)$
plays a special role in the quantification of chaotic and regular
processes due to maximal information about the effects of
Markovity and non-Markovity, short-range and long-range memory,
randomness and regularity effects. For the X-ray emission
intensity of the microquasar we have $\varepsilon_1(0)=4.94$.}
\end{figure}

The frequency dependence of the first three points of the
non-Markovity parameter $\varepsilon_{i}$, where $i=1,2,3 $ for
the intensity of the X-ray emission of the microquasar GRS
1915+105 is presented in Fig. 8. Of special value is the first
point of the non-Markovity parameter on zero frequency
$\varepsilon_{1}(0)=4.94$. This value occupies an intermediate
position between the appropriate values for a gas and a liquid.
For all the frequency dependencies in the region of frequencies
$\nu$=$0.057\div0.072 f.u.$ a series of peaks is detected.

In Table 1 some kinetic ($\lambda_1, \lambda_2, \lambda_3$) and
relaxation parameters ($\Lambda_1, \Lambda_2, \Lambda_3$) for the
X-ray intensity of GRS 1915+105 with a constant time step are
submitted. Let us note, that the kinetic parameter $|\lambda_1|$
means relaxation rate of the studied system. Small values of
kinetic and relaxation parameters signify the manifestation of
stochastic effects, randomness and stochastic behavior in the
registered time signals of the X-ray emission intensity of the
microquasar GRS 1915+105.

\begin{flushleft}
\small
 Table 1. {Some kinetic and relaxation parameters (absolute values) for
the X-ray intensity of GRS 1915+105 (constant time step)}
\end{flushleft}
\begin{center}
\scriptsize
\begin{tabular}{|p{1cm}|p{1cm}|p{1cm}|p{1cm}|p{1cm}|p{1cm}|}
\hline
$\lambda_{1}~ [\tau^{-1}]$& $\lambda_{2}~ [\tau^{-1}]$&
$\lambda_{3}~ [\tau^{-1}]$& $\Lambda_{1}~ [\tau^{-2}]$&
$\Lambda_{2}~ [\tau^{-2}]$& $\Lambda_{3}~ [\tau^{-2}]$ \\
\hline \hline
0.22& 1.20& 1.04& 0.10& 0.10& 0.03 \\
\hline
\end{tabular}
\bigskip \end{center}
\normalsize

\subsection{The analysis of the X-ray emission intensity
of the microquasar GRS 1915+105 for a time series with a variable
step}

The initial record of the X-ray emission intensity of the
microquasar GRS 1915+105 as a sequence of events is submitted in
Fig. 9. The given series essentially differs from the time series
with a constant time step (the registration was carried out in the
same period) by a visibly big set of the experimental data and
consequently is more informative. It allows to define quantitative
and qualitative parameters and properties of the studied system
with a higher degree of accuracy and reliability.

\begin{figure}[ht!]
     \leavevmode
\centering
\includegraphics[width=3.3in, height=2.5in, angle=0]{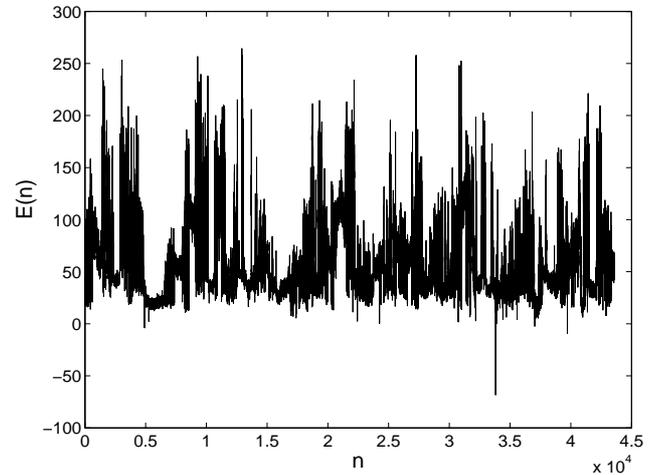}
\caption{The discrete sequence of events in the X-ray emission
intensity of the microquasar GRS 1915+105 (time series with
variable time step, mean time discretization is equal to 96 min).
Each point in the given sequence represents a single event of
emission.}
\end{figure}

The power spectra of the initial ECF (Fig. 10а) and three memory
functions of junior orders (Figs. 10b-d) are shown in Fig. 10. All
the figures are submitted on a double logarithmic scale. On the
whole frequency region of the power spectrum $\mu_{0}(\nu$) strong
fractality with the exponent $\alpha=-1.53$ (see Fig. 10a) is
observed. Let us note that the fractal parameter for the series
with a constant time step corresponds to $\alpha=-1.41$. Hence the
initial time signal registered with a variable time step is
characterized by greater randomness and stochastic behavior in
comparison with a signal with constant time step. Memory functions
$\mu_{i}(\nu)$, where $i=1,2,3$ (see Figs. 10b-d) manifest the
similar fractal behavior in a wide frequency interval.

\begin{figure}[ht!]
     \leavevmode
\centering
\includegraphics[width=3.3in, height=2.5in, angle=0]{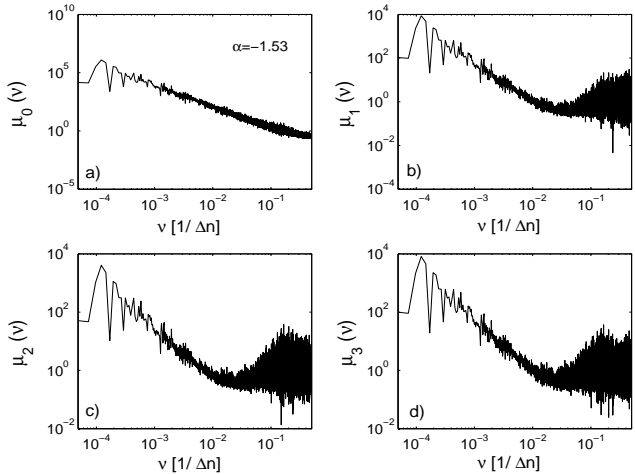}
\caption{The power spectrum of the initial correlation function
(a) and memory functions of the junior orders (b-d) of the X-ray
emission intensity of the microquasar GRS 1915+105 for a
non-equidistant time record. The power spectrum of the initial ECF
for the event series is characterized by greater fractality than
the power spectrum of TCF for the case of a constant time step. In
the power spectra of the memory function is observed the peak on
frequency $\nu=0.2 f.u.$ It reflects also the comparative analysis
of the fractal parameter for both cases.}
\end{figure}

In Fig. 11 the frequency dependencies of first three points of the
non-Markovity parameter $\varepsilon_{i}(\nu)$ (where $i=1,2,3$)
are submitted. The first point of the non-Markovity parameter
$\varepsilon_{1}(\nu)$ in the event presentation (see Fig. 11a)
allows to reveal additional features in relaxation processes of
the X-ray emission intensity of the microquasar GRS 1915+105. The
value of the parameter on zero frequency has increased more than
twice and is equal to $\varepsilon_{1}(0)=11.65$. The similar
behavior of the non-Markovity parameter (for the series with a
variable time step) is connected with the amplification of
quasi-Markovian and stochastic effects in the X-ray emission
processes. In Fig. 11 markovization effects are observed in
frequency behavior $\varepsilon_{1}(\nu)$ as a peak on frequency
$\nu=0.238$ $f.u$ (where $1 f.u=1/\Delta n$, $\Delta n=$1 event).
The values of the two subsequent points of the non-Markovity
parameter (see Fig. 11b, c) are close to a unit and are of
identical kind for both time series.

\begin{figure}[ht!]
     \leavevmode
\centering
\includegraphics[width=3.3in, height=1.6in, angle=0]{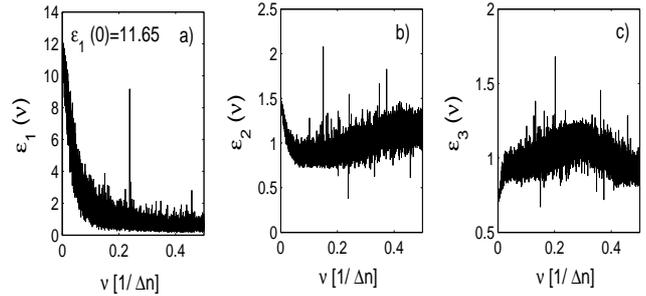}
\caption{The frequency dependence of the first three points of the
non-Markovity parameter of the X-ray emission intensity of the
microquasar GRS 1915+105 for an event correlation function (for
series of events presented in Fig. 9). The first point of the
non-Markovity parameter $\varepsilon_{1}(\nu)$ (a) is
characterized by the following features: the value on zero
frequency is equal to $\varepsilon_{1}(0)=11.65$; the peak on
frequency $\nu=0.238$ $f.u.$ connected with the sharp
amplification of Markovian (stochastic) effects is observed.}
\end{figure}

In Table 2 some kinetic ($\lambda_1, \lambda_2, \lambda_3$) and
relaxation parameters ($\Lambda_1, \Lambda_2, \Lambda_3$) for the
X-ray intensity of GRS 1915+105 with a variable time step are
submitted. The parameter $ | \lambda_1 | $ it is equal to 0.15.
Smaller rate of a relaxation for the time series with variable
time step is connected to more obvious and more pronounced
stochastic behavior. It may be, that the system needs more time
for return to a stable state of equilibrium.

\begin{flushleft}
\small Table 2. {Some kinetic and relaxation parameters (absolute
values) for the X-ray intensity of GRS 1915+105 (variable time
step)}
\end{flushleft}
\begin{center}
\scriptsize
\begin{tabular}{|p{1cm}|p{1cm}|p{1cm}|p{1cm}|p{1cm}|p{1cm}|}
\hline $\lambda_{1}~ [\tau^{-1}]$& $\lambda_{2}~ [\tau^{-1}]$&
$\lambda_{3}~ [\tau^{-1}]$& $\Lambda_{1}~ [\tau^{-2}]$&
$\Lambda_{2}~ [\tau^{-2}]$& $\Lambda_{3}~ [\tau^{-2}]$ \\
\hline \hline
0.15& 1.16& 1.03& 0.05& 0.12& 0.02 \\
\hline
\end{tabular}
\bigskip \end{center}
\normalsize

Thus, the approach to the description of statistical discrete
processes in the event presentation submitted in this work allows
to received more detailed and clear picture of the stochastic
processes occurring in complex nature systems.

\section{Discussion and Conclusion}

Energetic and spectral properties, quasi-periodic oscillations
(QPO) and the diverse temporal variability of GRS 1915+105 are the
focus of numerous studies in last years.

In the paper \cite{Vio} the problem of the limits concerning the
physical information that can be extracted from the analysis of
one or more time series typical of astrophysical objects has been
considered. The work \cite{Klein-Wolf} has clearly established
that radio emission from GRS 1915+105 is intimately related to the
presence of hard (power-law dominated) intervals in the X-ray
light curves. This in turn physically implies a clear relation
between a radiatively inefficient flow close to the black hole,
and a synchrotron-emitting overflow or jet. This suggest that MHD
effects could be responsible for the production and/or confinement
of the jets found in system. For the first time in Ref.
\cite{Fuchs} were observed concurrently in GRS 1915+105 all of the
following properties: a strong steady optically thick radio
emission corresponding to a powerful compact jet resolved with the
VLBA, bright near-IR emission, a strong QPO  at $2.5 Hz$ in the
X-rays and a power law dominated spectrum without any cuttof in
the $3-400 keV$ range.

J. Rodriguez et al. in the work \cite{Rodriguez} have presented
the results of simulations INTEGRAL and RXTE observations of the
microquasar GRS 1915+105. They have focused on the analysis of the
unique highly variable observation and have shown that they might
have observed a new class of variability. Then they have studied
the energetic dependence of a low frequency QPO from steady
observations.

A scenario for the variability of the microquasar GRS 1915+105
starts from previous works, leading to the tentative
identification of the accretion-ejection instability as the source
of the low frequency QPO and other accreting sources
\cite{Tagger}. A model for the $\sim30$ minute cycles often
exhibited by GRS 1915+105 is determined by the advection of
poloidal magnetic flux to the inner region of the disk, and its
destruction by reconnection (leading to relativistic ejections)
with the magnetic flux trapped in the vicinity of the central
source. This could be extrapolated further to understand the
long-term variability of this and other microquasars.

In the paper \cite{Rodriguez2} authors discuss the possible origin
of the following behavior: the QPO spectra are well modelled with
a cut-off power law except on one occasion where a single power
law gives a satisfactory fit (with no cut-off at least up to $\sim
40 keV$). The cut-off energy evolves significantly from one
observations to the other, from a value of $\sim21.8 keV$ to
$\sim30 keV$ in the other observations where it is detected. It
was suggested in the work \cite{Rodriguez2} that the compact jet
detected in the radio contributes to the hard X-ray ($\geq20 keV$)
mostly through synchrotron emission, whereas the X-ray emitted
below $20 keV$ would originate through inverse Compton scattering.
The dependence of the QPO amplitude on the energy can be
understood if the modulation of the X-ray flux is contained in the
Componized photons and not in the synchrotron ones.

The variability pattern is characterized in Ref.
\cite{Hannikainen} by a pulsing behavior, consisting of a main
pulse and a shorter, softer, and smaller amplitude precursor
pulse, on a timescale of 5 minutes in the JEM-X $3-35 keV$ light
curve. It was revealed, that the rising phase is shorter and
harder than the declining phase, which is opposite to what has
been observed in other otherwise similar variability classes in
this source. The fit show the source to be in a soft state
characterized by a strong disc component below $\sim6 keV$ and
Comptonization by both thermal and non-thermal electrons at higher
energies.

The source, which was observed 3 times in the plateau state,
before and after a major radio and X-ray flare, showed strong
steady optically thick radio emission corresponding to powerful
compact jet resolved in the radio emission corresponding to
powerful compact jet resolved in the radio images, bright
near-infrared emission, a strong QPO at $2.5 Hz$ in the X-rays and
a powerful law dominated spectrum without cut-off in the $3-300
keV$ range \cite{Fuchs2}.

Relativistic jets are now in Ref. \cite{Corbel} believed to be a
fairly ubiquitous property of accreting compact objects, and are
intimately coupled with the accretion history. Associated with
rapid changes in the accretion states of the binary systems,
ejections of relativistic plasma can be observed at radio
frequencies on timescale of weeks before becoming undetectable.
However, recent observations point to long-term  effects of these
ejecta on the interstellar medium with the formation of large
scale relativistic jets around binary systems.

In Ref. \cite{Muno} authors by \emph{Rossi X-ray Timing Explorer}
have found that as the radio emission becomes brighter and
optically thick, the frequency of a ubiquitous $0.5-10 Hz$ QPO
decreases, the Fourier phase lags between hard ($11.5-60 keV$) and
soft ($2-4.3 keV$) in the frequency range of $0.01-10 Hz$ change
sign from negative to positive, the coherence between hard and
soft photons at low frequencies decreases, and the relativistic
amount of low-frequency power in hard photons compared to soft
photons decreases.

Energetic dependence of a low frequency QPO in GRS 1915+105 have
been analyzed in the work \cite{Rodriguez3}. The results presented
could find an explanation in the context of the Accretion-Ejection
Instability, which could appear as a rotating spiral or hot point
located in the disk, between its innermost edge and the
co-rotation radius.

In the present work the dynamic features in the behavior of a
particle in the given cell in condensed matter and the X-ray
emission intensity of the microquasar GRS 1915+105 were considered
jointly on the basis of the statistical theory of non-Markov
processes. For this purpose we used two various applications of
the submitted theory as it allows to estimate chaotic and regular,
random and stochastic, Markov and non-Markov processes defining
the dynamics of the studied system. By taking into account the
effects of discreteness, long-range and short-range memory,
statistical relaxation now we are able to define a set of valuable
parameters and characteristics, which contain detailed information
about the properties of the studied systems connected with
randomness.

On the basis of the comparative analysis of the initial time
series, phase portraits, power spectra of the initial TCF and the
first point of the non-Markovity parameter we have found some
typical features of the behavior of regular and chaotic, Markov
and non-Markov components of stochastic processes in studied
systems. The systems with high level of chaotization (a low
density gas, a dense gas) are characterized by weak correlation
between the initial signals, the presence of complex chaotic
closed structures in the phase portraits, the high value of the
fractal parameter (according to its absolute value) of the power
spectrum of TCF and the big value of the first point of the
non-Markovity parameter on zero frequency. In the similar systems
the effects of short-range or instantaneous statistical memory
that correspond to Markovian processes appear. In regular
stochastic processes strong correlations in the initial time
signals, clear symmetric structures of phase clouds are observed.
The last one consists of centralized nucleus of high
concentration. Here low values of the fractal parameter (according
to its absolute value) of the power spectrum of the initial TCF,
small values of the first point of the non-Markovity parameter on
zero frequency are observed. In the similar systems the effects of
long-range statistical memory can be seen, and relaxation time
scales of TCF and memory functions are comparable. The processes
describing the given systems are non-Markov ones.

The use of the above submitted technique allows to estimate
non-Markov effects, robustness and regularity, stochastic behavior
and randomness of the X-ray emission intensity of the microquasar
GRS 1915+105 when the initial signal with constant time step is
registered. Here the scale of time fluctuations, long-range
effects of memory, discreteness of various processes and states,
effects of dynamic alternation in the intensity of the X-ray
emission play an important role. In the dynamics of the initial
time signal of the X-ray emission intensity the fast change of
various regimes, sharp and unexpected alternation of various types
of fluctuations and correlations were observed. Taking into
account discreteness of the experimental data, statistical effects
of long-range memory and the constructive role of fluctuations and
correlations we have received a detailed information about the
properties and parameters which characterize statistical
properties of the fluctuating X-ray emission of microquasar GRS
1915+105.

Thus, in this work, a new approach to the description of discrete
non-Markov stochastic processes with a variable time step in the
event representation is presented. This approach is based on the
consecutive use of the idea suggesting the existence of the event
correlation functions. The similar correlation functions are new
physical quantities determining probabilistic interrelation
between a sequence of events. When analyzing the time signal (the
X-ray emission intensity of the microquasar GRS 1915+105) with a
variable time step the following characteristics were determined:
the power spectra of the initial correlation function and memory
functions of the junior orders, first three points of the
non-Markovity parameter, the fractal parameter of the power
spectrum of ECF and the value of the first point of the
non-Markovity parameter on zero frequency. These dependencies and
parameters allow to receive additional information about the
properties and characteristics of the studied system: the
quasi-Markov character of relaxation processes in the intensity of
the X-ray emission, amplification of Markov effects on certain
frequencies, the fractal character of the power spectra of ECF and
memory functions.

Finally, this paper only takes the first step in introducing the
concept of event correlation analysis of time series and defining
it in terms of quantities that can be calculated from an
experimental data. We believe, the method developed can form a
basis to start formulating further meaningful questions regarding
the notions and presentations for real complex systems.

\section{Acknowledgments}
The authors are grateful to Dr. F.M. Gafarov for the program
package and Dr. L.O. Svirina for technical assistance. This work
was partially supported by the RFBR (Grants no. 05-02-16639-a),
Grant of Federal Agency of Education of Ministry of Education and
Science of Russian Federation. This work has been supported in
part (P.H.) by the German Research Foundation, SFB-486, project
A10.

\end{document}